\documentstyle[12pt]{article}
\begin{document}
\baselineskip=5.5mm
\begin{titlepage}
%%%%%%%%%%%%%%%%%%%%%%%%%%%%%%%%%%%%%%%%%%%%%%%%%
\begin{flushright} 
KEK-TH-733  \\ 
KOBE-TH-00-10 
%KOBE-TH-98-02\\
\end{flushright}
%%%%%%%%%%%%%%%%%%%%%%%%%%%%%%%%%%%%%%%%%%%%%%%%%%%
\vspace{2.3cm}
\centerline{{\large{\bf Pseudo-Dirac Scenario}}}
\centerline{{\large{\bf for}}}
\centerline{{\large{\bf Neutrino Oscillations}}}
\par
\par
\par\bigskip
\par\bigskip
\par\bigskip
\par\bigskip
\par\bigskip
\renewcommand{\thefootnote}{\fnsymbol{footnote}} 
\centerline{{\bf Makoto Kobayashi}$^{(a)}$ 
%\footnote[1]{e-mail:kobayath@post.kek.jp} 
and {\bf C.S. Lim}$^{(b)}$ } 
%\footnote[2]{e-mail:lim@phys.kobe-u.ac.jp}}
\par
\par\bigskip
\par\bigskip
\centerline{$^{(a)}$ Institute for Particle and Nuclear Studies, KEK, Oho 1-1, 
Tsukuba 305-0801, Japan} 
\centerline{$^{(b)}$ Department of Physics, Kobe University, Nada, 
Kobe 657-8501, Japan}
%\centerline{{\bf C.S. Lim}\footnote[1]{e-mail:lim@phys.sci.kobe-u.ac.jp}}
\par
\par\bigskip
\par\bigskip
\par\bigskip
\par\bigskip
\par\bigskip
%\centerline{\today }\par
\par\bigskip
\par\bigskip
\par\bigskip
\par\bigskip
\centerline{{\bf Abstract}}\par
We argue how pseudo-Dirac scenario for neutrinos leads to rich 
neutrino oscillation phenomena, including oscillation inside each generation. 
The pseudo-Dirac scenario is generalized by incorporating generation mixings and formulae for the various neutrino oscillations are derived. As the application we compare the formulae with the corresponding data. We find that observed pattern of mixings, such as almost maximal mixing in the atmospheric neutrino oscillation, is naturally explained in the generalized Pseudo-Dirac scenario with small generation mixings. We, however, also point out that there remain some problems to be settled for this scenario to be viable. The possible theoretical framework to realize the pseudo-Dirac scenario is also briefly commented on.

\par\bigskip
\par\bigskip
\par\bigskip
\par\bigskip
\par\bigskip
\par\bigskip
\par\bigskip

%\noindent September 1995
\end{titlepage}
\newpage

\newcommand{\mass}{m_{H}}

\vspace{0.5 cm}
\leftline{\bf 1. Introduction}
\vspace{0.2 cm}
 
The presence of neutrino oscillation, strongly suggested by the recent Superkamiokande results 
on atmospheric neutrinos \cite{Kamatm}, is almost unique clue to  physics beyond the 
well-established 
standard model. More precisely, recent data  on neutrino oscillations seem to have put forward the following challenging theoretical problems, which may lead to the physics beyond the  standard model; \\ 
(a) The data from (Super-)Kamiokande on atmospheric neutrinos necessitate a large or 
almost maximal mixing angle \cite{Kamatm}. Solar neutrino deficit may also be explained by the ``large angle 
solution" \cite{Kamsolar}. How can such large or maximal mixing be naturally 
derived theoretically ? \\ 
(b) If we further accept LSND result \cite{LSND}, in addition to the solar and atmospheric neutrino data, the scheme with only 3 light neutrino states clearly gets into trouble.  
What kind of theoretical framework or model is needed to accommodate all of these neutrino oscillations ? It has been argued that we should introduce at least one ``sterile state". \\ 
(c) There seems to be a large disparity among the magnitudes of mixing angles implied by these experiments; 
 The data on atmospheric neutrino and possibly that on solar neutrino indicate the necessity 
of large or almost maximal mixing angles \cite{Kamatm}, \cite{Kamsolar}, while the neutrino oscillation at LSND experiment is well described by   a  small mixing angle \cite{LSND}. How can such disparity be naturally explained theoretically ?

One clear thing is that these problems altogether suggest that flavour mixing 
or mass matrices in leptonic sector are quite different from those in quark sector. In particular the presence of large mixings and the necessity of extending the scheme with only 3 light neutrinos are  specific new features in the leptonic sector, not shared by the quark sector, and may lead to a drastic 
modification of the Standard Model. 

The main purpose of the present work is to generalize the pseudo-Dirac scenario for 
neutrino masses, which we argue to imply rich neutrino oscillation phenomena, 
by incorporating generation mixings and to derive general formulae for neutrino oscillations in the generalized pseudo-Dirac scenario. We will show that the above problems, (a), (b) and (c), are naturally (without any fine-tuning) solved simultaneously in the generalized scheme without enlarging the number of generations. It is worth noticing that small generation mixings are shown to be exactly what we need to solve the problems. We, however, will also point out some 
serious problems encountered by the atmospheric and solar neutrino oscillations into sterile states \cite{Kamatm2}, \cite{Kamsolar} and try to discuss some possible ways to cure  these problems. We also investigate briefly what kind of model or theoretical framework is possible for the pseudo-Dirac scenario to be realized, for the purpose of searching for some direction to the physics beyond the standard model.  
The key ingredient for our scenario is ``pseudo-Dirac" property of neutrinos, whose precise meaning is now discussed in some detail. 

It will be natural to suspect that the specific features of flavour mixing 
or mass matrices in leptonic sector stated above should be related with the peculiarity of the leptonic masses, i.e. the fact that only neutrinos may have Majorana  masses. 
Once neutrinos are allowed to have 
Majorana masses, we may think of three typical cases for neutrino masses, which we will discuss successively below. In the base of weak-eigenstates, $\psi_{wL}$,  where active states are put in the up-stairs and ``sterile" states are put in the down-stairs, 
\begin{equation}
\psi_{wL} = \pmatrix{\nu_{\alpha L} \cr
         \overline{\nu}_{\alpha L}} \ \ \ \  
         (\alpha = e, \mu, \tau; \ \ \overline{\nu}_{\alpha L} = (\nu_{\alpha R})^C ), 
\end{equation} 
the neutrino mass term is generally written as 
\begin{equation}
L_{\mbox{mass}} = \frac{1}{2}\psi_{wL}^t C M \psi_{wL}, 
\end{equation} 
where $C$ is the charge-conjugation matrix and the $6 \times 6$ mass matrix $M$  takes  a form of 
\begin{equation}
M = \pmatrix{M_L & M_D^t \cr
             M_D & M_R^{\ast}}, 
\end{equation} 
with the $3 \times 3$ matrices $M_D$, $M_L$ and $M_R$ being those for Dirac 
masses, left- and right-handed Majorana masses, respectively.  
Depending on the extent of lepton number violation, or relative magnitudes of Majorana masses to those of Dirac masses, we can 
think of three typical cases. 

\noindent (1) pure Dirac \\ 
Imposing lepton number conservation, or ignoring all Majorana masses ($M_L = M_R = 0$), we get pure Dirac neutrinos. We all know 
that there are only 3 mass-eigenvalues for 3 generations, although the mass matrix $M$ 
should have 6 eigenvalues, in general. What's really happening is that there are 3 degenerate pairs of mass-eigenstates. The situation may be easily seen for simplified 1 generation case 
\begin{equation}
M = \pmatrix{0 & m_D \cr
             m_D & 0}. 
\end{equation} 
It is easy to see that $M$ has eigenvalues $m_D, -m_D$ and the angle, $\theta$, in the orthogonal matrix to diagonalize $M$ is just $\pi /4$. If processes we are interested in have no chirality flip, as in the case of neutrino oscillations (in the absence of magnetic field), the sign of mass is irrelevant and we have degenerate mass-squared. It should also be noted that in this case a maximal mixing, $\theta = \pi /4$, between an active state and a sterile state, $\nu_L$ and $\overline{\nu}_L$, has been realized. Unfortunately, this maximal mixing does not lead to any neutrino oscillation, just because the mass-squared are 
degenerated. 

\noindent (2) ``Pseudo-Dirac"  \\ 
What happens if we allow small lepton number violation, i.e. if we switch on very small Majorana masses, $M_L, M_R \ll M_D$ 
(with the magnitudes of matrices being compared by taking typical orders of 
magnitudes of the matrix elements). Neutrinos are still almost Dirac particles and are called  ``pseudo-Dirac" neutrinos \cite{Wolfenstein}. The small Majorana masses, however, slightly lift the degeneracy of mass-eigenvalues, and we get almost degenerate pairs of eigenstates with tiny mass differences. As far as the Majorana masses are small, the mixing angles should remain almost maximal, $\theta 
\simeq \pi /4$. 
To understand the situation, we consider the 1 generation case again. The mass matrix now reads as 
\begin{equation}
M = \pmatrix{m_L & m_D \cr
             m_D & m_R},  
\end{equation}
with $m_L, m_R \ll m_D $. We now get $|\mbox{tan} 2\theta| = |\frac{2m_D}{m_R - m_L}| \gg 1$, leading to almost maximal mixing $\theta \simeq \frac{\pi}{4}$. We have two mass eigenstates, which are almost symmetric and anti-symmetric combinations of active and sterile states, i.e., $\nu_{S} = \mbox{sin} \theta \ \nu_{L} + \mbox{cos} \theta \ \overline{\nu}_{L} \simeq \frac{1}{\sqrt{2}}(\nu_{L} + \overline{\nu}_{L})$, and $\nu_{A} = (-i) (\mbox{cos} \theta \ \nu_{L} - \mbox{sin} \theta \ \overline{\nu}_{L}) \simeq \frac{1}{\sqrt{2}i}(\nu_{L} - \overline{\nu}_{L})$. Their masses are almost degenerate but are slightly different; $m_S \sim m_A \sim m_D$, $\Delta m \equiv 
m_S - m_A \sim m_L + m_R \ll m_D$. Now the tiny mass difference and the almost maximal mixing will lead to a neutrino oscillation between an active state and a sterile state, even if we have only 1 generation ! 
Actually even for the 1 generation case two kinds of neutrino oscillations, without and with chirality flip,  are possible (though the latter oscillation 
necessitates the presence of magnetic field and we ignore the magnetic field unless otherwise stated@in this paper), i.e. (i) $\nu_L \rightarrow \overline{\nu}_L$ and (ii) $\nu_L \rightarrow \nu_R$ \cite{Kobayashi}.    If matter effects are included, the oscillations of (i) and (ii) become resonant oscillations, which are quite similar to those in MSW \cite{MSW} and Resonant Spin Flavour Precession (RSFP) \cite{RSFP} scenarios: The Hamiltonian in the base of $(\nu_L, \overline{\nu}_L )$ for the former resonant oscillation is given by \cite{Kobayashi} 
\begin{equation}
H = \pmatrix{a & \frac{\Delta m^2}{4E} \mbox{sin} 2\theta \cr
             \frac{\Delta m^2}{4E} \mbox{sin} 2\theta & \frac{\Delta m^2}{2E} \mbox{cos} 2\theta},   
\end{equation}
where the matter effect $a$ is given for e.g. $\nu_{e}$ as $a_{e} = \frac{G_F}{\sqrt{2}}(2N_e - N_n)$, with $N_e$ and $N_n$ denoting the number densities of electron and neutron, respectively.  

\noindent (3) See-saw  \\ 
The last possibility is famous see-saw scenario \cite{Seesaw} in which SU(2) invariant Majorana masses, $M_R$, are supposed to be much larger than the Dirac masses: 
$M_R \gg M_D, \ M_L \simeq 0$. The sterile states 
$\overline{\nu}_L$ approximately become mass-eigenstates and are decoupled from low energy processes such as neutrino oscillation. Thus only lighter 3 mass-eigenstates ($\simeq \nu_{\alpha L}$) participate in neutrino oscillation phenomena. In the see-saw scenario, therefore, the mixings relevant for the neutrino oscillations are generation mixings and there seems to be no immediate reason to expect large mixing angles. It is also worth noting that as far as chirality preserving oscillations are concerned, there is no observable distinction between the cases of (1) and (3); In both cases only 3 light neutrino states participate in the oscillations. 

From the above discussion we learn that only in the pseudo-Dirac scenario 
6 neutrino states fully participate in low energy processes, and rich neutrino 
oscillation phenomena, both inter-generational and active $\leftrightarrow $ sterile, are expected. In fact a trial to explain existing data on neutrino oscillations based on the pseudo-Dirac scenario was made some time ago 
\cite{Kobayashi}, \cite{Minakata}. (For the recent revived interest in this scenario, refer to \cite{Geiser}, \cite{Chang}. See also Ref.\cite{Yasuda} for the recent discussions in the framework of four neutrinos with only one sterile state.)
 In our previous attempt \cite{Kobayashi} generation mixings were switched off, for brevity, and three generations shared their roles to account for neutrino oscillations; Solar neutrino oscillation was mainly due to 
$\nu_{eL} \rightarrow \overline{\nu}_{eL}$ in the first generation (in  ref.\cite{Kobayashi} the effect of magnetic field was also taken into account). The atmospheric neutrino oscillation could be naturally explained by $\nu_{\mu L} \rightarrow \overline{\nu}_{\mu L}$ with almost maximal mixing angle. At that time there was a datum to suggest the existence of 17KeV neutrino \cite{Simpson}, which enforced us to rely on the pseudo-Dirac scenario, as otherwise three generation scheme could not explain all of these data simultaneously. The 17 KeV neutrino has been ruled out, and instead  there has appeared the LSND data \cite{LSND}, which again necessitates three independent mass differences and to modify the ordinary 3 generation scheme with either see-saw or pure Dirac neutrinos.

In the present paper we generalize our previous argument \cite{Kobayashi} including generation mixings. We will find that small generation mixings, just as in the CKM matrix in the quark sector,  is exactly what we need: the nice features of pseudo-Dirac scenario, such as maximal mixings in atmospheric and solar neutrino oscillations, are known to 
remain basically intact, while LSND data is naturally explained by the small generation mixings. In such a sense, we may say that \\ 
{\it ``The recent data on neutrino oscillations may be natural consequences of the property that lepton number is only slightly violated and generation mixings are small."}

\vspace{0.5 cm}
\leftline{\bf 2. Mass eigenstates and mixings}
\vspace{0.2 cm}

In the present work we allow arbitrary generation mixings or arbitrary off-diagonal mass matrices, except for the assumption of pseudo-Dirac property $M_L, M_R \ll M_D$. 
Thus we may naively expect that diagonalization of $6 \times 6$ mass matrix is quite complicated and various formulae for the probabilities of neutrino oscillations are expressed by use of 6 mass-eigenvalues and 15 
mixing angles (together with possible multiple CP violating phases). It, however, turns out that under the assumption of pseudo-Dirac, all probabilities of neutrino oscillations are describable in terms of 6 mass eigenvalues and just one $3 \times 3$ unitary matrix $U$, ``MNS matrix" \cite{MNS}, which has three mixing angles and one CP violating phase and just corresponds to the CKM matrix in quark sector.   

To see this let us now discuss the diagonalization of the mass matrix $M$. 
Ignoring magnetic field, only chirality preserving transitions are important. So what we should diagonalize is $M^{\dagger}M$, rather than $M$ itself. Keeping 
terms up to the first order in Majorana masses, which may be justified in the pseudo-Dirac hypothesis, $M^{\dagger}M$ reads as 
\begin{equation}
M^{\dagger}M \simeq  
 \pmatrix{ M_D^{\dagger} M_D & M_L^{\ast}M_D^{t} + M_D^{\dagger}M_R^{\ast} \cr
           M_D^{\ast} M_L + M_R M_D & M_D^{\ast} M_D^{t}}. 
\end{equation}
Because of pseudo-Dirac property, the dominant matrix is $M_D$. 
Thus we first diagonalize it by bi-unitary transformation;  
\[
U_R^{\dagger} \ M_D \ U_L = \mbox{diag} (m_{1}, m_{2}, m_{3})  
\equiv \hat{M},   
\] 
\begin{equation}
\nu_{\alpha L} = (U_L)_{\alpha i} \ \nu_{iL},  \ \ \ 
\overline{\nu}_{\alpha L} = (U_R^{\ast})_{\alpha i} \ \overline{\nu}_{iL}.
\end{equation}
Accordingly, $M^{\dagger}M$ is cast into the following form by a unitary transformation due to a $6 \times 6$ unitary matrix $V$,
\[
V^{\dagger} \ (M^{\dagger} \ M) \ V  = 
\pmatrix{ \hat{M}^2 & U_L^{\dagger} M_L^{\dagger} U_L^{\ast} \hat{M} +  \hat{M} U_R^{\dagger}M_R^{\ast} U_R^{\ast} 
          \cr 
    \hat{M} U_L^{t} M_L U_L + U_R^{t} M_R U_R \hat{M} &  \hat{M}^2},  
\]  
\begin{equation} 
V = \pmatrix{U & 0 \cr
             0 & U_R^{\ast}},  \ \ \ U \equiv U_L,  
\end{equation}
where  $U_L$ has been rewritten simply as $U$, as it is the only matrix which appears in the formulae of neutrino oscillations. $\hat{M}^2$ is a diagonal matrix, while the matrices in off-diagonal position, e.g. $\hat{M} U_L^{t}M_L U_L + 
U_R^{t} M_R U_R \hat{M}$, have not been diagonalized yet. Such off-diagonal matrices, however, are much smaller than $\hat{M}^2$ 
 due to the pseudo-Dirac property and seem to be negligible, anyway. It is not quite right, since 
 once we ignore these off-diagonal matrices there appear degenerate pairs in the eigenvalues of $M^{\dagger} M$, i.e., each of $m_i^2 \ (i = 1,2,3)$ appear twice. It is a general wisdom in perturbation theory that when there is a degenaracy in eigenvalues, eigenvalues and 
eigenstates can be fixed only after we include first order perturbation, that connects the members of the pair, while other perturbations connecting different pairs may be safely ignored. This means that $V^{\dagger} \ (M^{\dagger} \ M) \ V$ can be effectively decomposed into three independent block-diagonal matrices. Each block-diagonal matrix takes a form of, 
\begin{equation} 
\pmatrix{m_i^2 & m_i \epsilon_{i}^{\ast} \cr 
         m_i \epsilon_{i} & m_i^2}  \ \ \ \ (i = 1,2,3),    
\end{equation}   
where $\epsilon_{i} \equiv (U_L^{t} M_L U_L + U_R^{t} M_R U_R)_{ii}$, 
and $|\epsilon_{i}| \ll m_i$, because of the pseudo-Dirac property. 

It is now easy to see that we obtain, in total,  6 mass eigenstates 
\begin{equation} 
\nu_{iS} \equiv \frac{1}{\sqrt{2}} (\nu_{iL} + e^{i \phi_{i}} \overline{\nu}_{iL}), \ \ \ 
\nu_{iA} \equiv \frac{1}{\sqrt{2}i} (\nu_{iL} - e^{i \phi_{i}} \overline{\nu}_{iL}),  \ \ \ (i =1,2,3),     
\end{equation} 
where $e^{i \phi_{i}} = \epsilon_{i} / |\epsilon_{i}|$. 
Their mass-eigenvalues are given as  
\begin{equation}
m^{2}_{iS} = m_i^2 + m_i |\epsilon_{i}|, \ \ \ m^{2}_{iA} = m_i^2 - m_i |\epsilon_{i}|, \ \ \ 
(i = 1,2,3). 
\end{equation}
To summarize, $\psi_{wL}$ is related to mass-eigenstates as,  
\begin{equation}
\psi_{wL} = 
 \pmatrix{\nu_{\alpha L} \cr 
         \overline{\nu}_{\alpha L}}
 = \hat{V} \ \pmatrix{\nu_{iS} \cr 
                      \nu_{iA}},  
\end{equation}
where 
\begin{equation}
\hat{V} \equiv \pmatrix{U & 0 \cr
             0 & U_R} \cdot 
           \pmatrix{\frac{1}{\sqrt{2}} & 0 & 0 & \frac{i}{\sqrt{2}} & 0 & 0 \cr
                    0 & \frac{1}{\sqrt{2}} & 0 & 0 & \frac{i}{\sqrt{2}} & 0 \cr                     0 & 0 & \frac{1}{\sqrt{2}} & 0 & 0 & \frac{i}{\sqrt{2}} \cr    \frac{1}{\sqrt{2}} e^{-i\phi_{1}} & 0 & 0 & -\frac{i}{\sqrt{2}} e^{-i\phi_{1}} & 0 & 0 \cr      
   0 & \frac{1}{\sqrt{2}} e^{-i\phi_{2}} & 0 & 0 & -\frac{i}{\sqrt{2}} e^{-i\phi_{2}} & 0 \cr      
   0 & 0 & \frac{1}{\sqrt{2}} e^{-i\phi_{3}} & 0 & 0 & -\frac{i}{\sqrt{2}} e^{-i\phi_{3}}
   },  
\end{equation}
and correspondingly the mass-squared matrix is diagonalized as 
\begin{equation} 
\hat{V}^{\dagger} \ (M^{\dagger} \ M) \ \hat{V} = M_{\mbox{diag}}^{2},  
 \ \ \ M_{\mbox{diag}} = \mbox{diag} (m_{1S}, m_{2S}, m_{3S}, m_{1A}, m_{2A}, m_{3A}). \end{equation} 
Now the neutrinos emitted by weak interactions (weak eigenstates) are expressed in terms of 
mass eigenstates $\nu_{jS}, \nu_{jA} \ (j=1-3)$ and a unitary matrix $U$, as follows: 
\begin{equation} 
\nu_{\alpha L} = U_{\alpha j} \frac{\nu_{jS} + i \nu_{jA}}{\sqrt{2}}. 
\end{equation} 
The fact that there appears only  single $3 \times 3$ unitary matrix $U$, even though we started from an arbitrary $6 \times 6$ 
 mass matrix $M$, is one of our main results based on the pseudo-Dirac 
 property.

\vspace{0.5 cm}
\leftline{\bf 3. Formulae for neutrino oscillations}
\vspace{0.2 cm}

We will now derive the formulae for neutrino oscillations in terms of the differences of 6 mass-squared and single unitary matrix $U$.  
Though there is no reason to expect  apriori some specific pattern of neutrino 
masses, we can still get some useful information on the pattern from the reported data on neutrino oscillations \cite{Kamatm}, \cite{Kamsolar}, \cite{LSND}. Namely, once we regard the mixing angles in $U$ as small, as suggested by the CKM matrix in quark sector,  the mass-squared difference, responsible for  each observed neutrino oscillation, is given as 
\[  
\mbox{solar neutrino}: m_1 |\epsilon_{1}| \sim 10^{-5} - 10^{-4} (eV^2), 
\] 
\[ 
\mbox{atmospheric neutrino}: m_2 |\epsilon_{2}| \sim 10^{-3} - 10^{-2} (eV^2),  \] 
\begin{equation}    
\mbox{LSND}: \Delta m_{12}^2 \sim 10^{-1} - 1 (eV^2).  
\end{equation} 
This knowledge suggests (with a little prejudice) a hierarchical structure of mass differences 
\begin{equation}  
m_1 |\epsilon_{1}| \ll m_2 |\epsilon_{2}| \ll m_3 |\epsilon_{3}| \ll 
\Delta m_{12}^2 \ll \Delta m_{13}^2,   
\end{equation}
where $\Delta m_{ij}^2 \equiv m_j^2 - m_i^2 $.
The hierarchical structure makes the formulae for neutrino oscillations simple  and easy to be compared with the data. 

\noindent 3.1 \ A general formula for vacuum oscillation 

We first note that except for the case of solar neutrino oscillation, both of 
atmospheric neutrino oscillation and the oscillation in LSND experiment are well described by vacuum oscillations. This is definitely true for LSND case, but may need some care in the case of atmospheric neutrino, as will be commented on 
below. 

In general, the probability of finding  a state, born as an active state $\nu_{\alpha L}$ at time 0, in an active state $\nu_{\beta L}$ ($\alpha$ and $\beta$ 
may be the same) at time $t$ is given by
\begin{equation} 
P(\nu_{\alpha} \rightarrow \nu_{\beta}) = |(\hat{V} \mbox{exp}\{i \frac{M_{\mbox{diag}}^{2}}{2E}t \} \hat{V}^{\dagger})_{\beta \alpha}|^2 
=  \frac{1}{4}|\sum_{j = 1}^{3}U_{\beta j} \{ \mbox{exp}(i \frac{m_{jS}^2}{2E} t ) + \mbox{exp}
(i \frac{m_{jA}^2}{2E} t) \} U^{\ast}_{\alpha j}|^2 . 
\end{equation}   

\noindent 3.2 \ Formulae for atmospheric neutrino oscillation  

As the oscillation of atmospheric neutrino is sensitive to the mass difference  
$ m_2 |\epsilon_{2}| \sim 10^{-3} \ (eV^2)$ \cite{Kamatm}, under the mass hierarchy eq.(18) 
$m_1 |\epsilon_{1}|$ may be ignored and $\nu_{1}$ can be 
regarded as pure Dirac particle, i.e. $\nu_{1} = \frac{1}{\sqrt{2}}\{ (\nu_{1S}  + c.c.) + i (\nu_{1A}  + c.c.) \}$  with a unique mass $m_{1S} = m_{1A} = m_{1}$. 
The oscillation of atmospheric neutrino is due to the interference between $\nu_{2S}$ and $\nu_{2A}$, and  the matter waves of other states, $\nu_{1}, \nu_{3S}$ and $\nu_{3A}$,  do not interfere with  $\nu_{2S}, \nu_{2A}$ or with each other, when time-average is taken for the high frequency modes in the oscillation probability. Thus the formula for the probability of atmospheric $\nu_{\mu}$ to survive till time $t$, relevant for the zenith angle distribution, simply reads as \begin{equation} 
P(\nu_{\mu L} \rightarrow \nu_{\mu L})_{\mbox{atm}} = |U_{\mu 1}|^4 + |U_{\mu 2}|^4 \mbox{cos}^2 (\frac{m_2 |\epsilon_2|}{2E} t) + \frac{1}{2} |U_{\mu 3}|^4 . 
\end{equation}
Let us note that there are constant terms $|U_{\mu 1}|^4, \  |U_{\mu 3}|^4$ coming from the time-average of the high frequency modes, in sharp contrast to the  conventional formula in a simplified 2 states system. 
The following formulae are also relevant for the analysis of atmospheric neutrino oscillation:   
\begin{eqnarray} 
P(\nu_{e L} \rightarrow \nu_{e L})_{\mbox{atm}} &=& |U_{e 1}|^4 + |U_{e 2}|^4 
\mbox{cos}^2 (\frac{m_2 |\epsilon_2|}{2E} t) + \frac{1}{2} |U_{e 3}|^4 ,  \\ 
P(\nu_{\mu L} \rightarrow \nu_{e L})_{\mbox{atm}} &=& |U_{\mu 1}|^2 |U_{e 1}|^2 + |U_{\mu 2}|^2 |U_{e 2}|^2 
\mbox{cos}^2 (\frac{m_2 |\epsilon_2|}{2E} t) + \frac{1}{2} |U_{\mu 3}|^2 |U_{e 3}|^2 .  
\end{eqnarray}

\noindent 3.3 \ Formula for LSND neutrino oscillation 

As the neutrino oscillation observed by LSND is sensitive to the mass difference  
$\Delta m_{12}^2 \sim 10^{-1} - 1 \ (eV^2)$ \cite{LSND}, under the hierarchy (18) all neutrino states can be 
regarded as pure Dirac particles, i.e. $m_i |\epsilon_{i}| = 0$ and $m_{iS}^2 = m_{iA}^2 = m_{i}^2 \ (i = 1,2,3)$. The LSND neutrino oscillation is, therefore,  due to the interference between $\nu_{1}$ and 
$\nu_{2}$, and  the matter waves of another state, $\nu_{3}$, does not interfere with $\nu_{1}, \nu_{2}$, when time-average is taken. Thus the formula for the 
transition probability of LSND neutrino simply reads as 
\begin{eqnarray}  
P(\nu_{\mu L} \rightarrow \nu_{e L})_{\mbox{LSND}} &=& |U_{\mu 1}|^2 |U_{e 1}|^2 + |U_{\mu 2}|^2 |U_{e 2}|^2 + |U_{\mu 3}|^2 |U_{e 3}|^2  \nonumber \\  
&+& 2 \mbox{Re}(U_{\mu 1} U_{e1}^{\ast} U_{\mu 2}^{\ast} U_{e2})
\mbox{cos} (\frac{\Delta m_{21}^2}{2E} t) -2 \mbox{Im} (U_{\mu 1} U_{e1}^{\ast} U_{\mu 2}^{\ast} U_{e2}) \mbox{sin} (\frac{\Delta m_{21}^2}{2E} t)  \nonumber \\ &=& 4\{ |U_{\mu 2} U_{e2}|^2 + \mbox{Re}(U_{\mu 2}^{\ast} U_{e2}U_{\mu 3} U_{e3}^{\ast}) \} \mbox{sin}^2 (\frac{\Delta m_{21}^2}{4E} t) \nonumber \\ 
 &+& 2 \mbox{Im}(U_{\mu 2}^{\ast} U_{e2}U_{\mu 3} U_{e3}^{\ast}) \mbox{sin} (\frac{\Delta m_{21}^2}{2E}t) + 2 |U_{\mu 3} U_{e3}|^2 .
\end{eqnarray}

\noindent 3.4 \ The matter oscillation of solar neutrino  

The time-evolution of the system in the presence of matter effects are governed by 
\begin{equation} 
i \frac{d}{dt}  \pmatrix{\nu_{\alpha L} \cr 
                             \overline{\nu}_{\alpha L}} 
  =  H  \pmatrix{\nu_{\alpha L} \cr 
                             \overline{\nu}_{\alpha L}},  
\end{equation} 
where 
\begin{eqnarray} 
 H &=&  \frac{1}{2E} (\hat{V}  M_{\mbox{diag}}^{2} \hat{V}^{\dagger} ) 
     +  A,  \nonumber \\ 
   &=&  \hat{V} \{ \frac{1}{2E} M_{\mbox{diag}}^{2} + 
   \hat{V}^{\dagger} A \hat{V} \} \hat{V}^{\dagger},  \ \ \ A \equiv 
   \mbox{diag} (a_{e}, a_{\mu}, a_{\tau}, 0, 0, 0),
\end{eqnarray}
and the elements of the matrix $A$ denoting the matter effects of $\nu_e, \nu_{\mu}, \nu_{\tau}$ are given as $a_{e} = \frac{G_F}{\sqrt{2}}(2N_e - N_n), \ a_{\mu} = a_{\tau} = \frac{G_F}{\sqrt{2}}(- N_n)$. 
It is easy to see that for the energy range of solar neutrinos and the mass hierarchy of (18), the matter effects inside the sun satisfy 
\begin{equation}
E \ a_{\alpha} \sim m_1 |\epsilon_{1}|\ll m_2 |\epsilon_{2}| \ll \ldots 
\ll \Delta m_{13}^{2}  \ \ (\alpha = e, \mu, \tau). 
\end{equation}
Under the hierarchical structure of mass differences and the matter effects, in  the Hamiltonian in the base of the mass eigenstates, $\frac{1}{2E} M_{\mbox{diag}}^{2} + \hat{V}^{\dagger} A \hat{V}$, all off diagonal matrix elements due to the matter effects  can be safely  ignored except the ones in 2 $\times$ 2 subsystem of $\nu_{1S}$ and  $\nu_{1A}$. Thus the heavier states,  $\nu_{iS}, \nu_{iA} \ (i = 2, 3)$ are decoupled from the two states subsystem of $(\nu_{1S}, \nu_{1A})$, and the time-evolution of the subsystem is governed by a  $2 \times 2$  Hamiltonian 
\begin{equation}
  \frac{1}{2E} \pmatrix{ m_{1S}^{2} & 0  \cr 
                           0        &  m_{1A}^{2} } 
   +  \frac{1}{2} (\sum_{\alpha = e}^{\tau}\ |U_{\alpha 1}|^2 a_{\alpha})
   \pmatrix{1 & 1  \cr 
            1 & 1 }. 
\end{equation} 
This Hamiltonian can be rewritten in the base of $(\nu_{1L}, \overline{\nu}_{1L})$ (ignoring a piece proportional to a unit matrix) as 
\begin{equation}
   \pmatrix{\sum_{\alpha = e}^{\tau} |U_{\alpha 1}|^2 a_{\alpha} & 
      \frac{\Delta m^2}{4E} \mbox{sin} \ 2\theta  \cr 
    \frac{\Delta m^2}{4E} \mbox{sin} \ 2\theta  & \frac{\Delta m^2}{2E} \mbox{cos} \ 2\theta },  
\end{equation}
with $\Delta m^2 = 2 m_1 |\epsilon_1|$ and $\theta = \frac{\pi}{4}$. This hamiltonian just corresponds to the one in the MSW mechanism \cite{MSW}, though in our case the mixing angle is maximal and the matter effect has been modified into  $ \sum_{\alpha = e}^{\tau} |U_{\alpha 1}|^2 a_{\alpha}$ . 
Now the survival probability of solar neutrino (time averaged) can be written in a 
simple form 
\begin{equation}
\overline{P}(\nu_{eL} \rightarrow \nu_{eL})_{\mbox{solar}}= |U_{e1}|^4 \ \overline{P}(\nu_{1L} \rightarrow \nu_{1L})_{\mbox{eff}} + \frac{1}{2} \ |U_{e2}|^4
                      + \frac{1}{2} \ |U_{e3}|^4 , 
\end{equation} 
where the survival probability $ \overline{P}(\nu_{1L} \rightarrow \nu_{1L})$ in the effective 2 states system of $(\nu_{1L}, \overline{\nu}_{1L})$ is calculable by use of  the Hamiltonian (28). 
A similar reduction formula based on a hierarchical mass structure was obtained in Ref.\cite{Lim} in order  to reduce solar neutrino oscillation in 3 generation model into  that in an effective 2 generation model. 

\vspace{0.5 cm}
\leftline{\bf 4. Pseudo-Dirac scenario confronted by the data on neutrino oscillations}
\vspace{0.2 cm}

\noindent 4.1 \ Comparing the formulae with the data 

As the application of the formulae we have derived for the neutrino oscillations,  we are now going to compare them with the corresponding experimental data. As we have already advertised we will see that small generation mixing angles in 
the unitary matrix $U$ is just what we need to explain the pattern of mixing angles observed in neutrino oscillations of our interest.

Thus we first consider the case where generation mixings are small, though the 
formulae we have derived above are applicable for arbitrary generation mixings. Retaining only the leading contributions for small generation mixing angles (say $\theta_{1}$, $\theta_{2}$, $\theta_{3}$ just as the angles in CKM matrix), we get the following formulae, relevant for each neutrino oscillation
\begin{eqnarray}
 \mbox{solar neutrino}&:& \ \ \ \overline{P}(\nu_{eL} \rightarrow \nu_{eL})_{\mbox{solar}} \simeq  \overline{P}(\nu_{1L} \rightarrow \nu_{1L})_{\mbox{eff}}, \\  
 \mbox{atmospheric neutrino}&:& \ \ \ 1 - P(\nu_{\mu L} \rightarrow \nu_{\mu L})_{\mbox{atm}} \simeq  \mbox{sin}^2 (\frac{m_2 |\epsilon_2|}{2E} t), \nonumber \\ 
                             & &  \ \ \  P(\nu_{eL} \rightarrow \nu_{eL})_{\mbox{atm}} 
                            \simeq  1, \nonumber \\ 
                             & &  \ \ \  P(\nu_{\mu L} \rightarrow \nu_{eL})_{\mbox{atm}} \simeq  0,   \\  
 \mbox{LSND}&:& \ \ \ P(\nu_{\mu L} \rightarrow \nu_{e L})_{\mbox{LSND}} 
\simeq  4|U_{e 2}|^2 \mbox{sin}^2 (\frac{\Delta m_{21}^2}{4E} t) . 
\end{eqnarray}
In the formula for LSND $|U_{e3}U_{\mu 3}|$ has been neglected compared with $|U_{e2}U_{\mu 2}|$, as is suggested by the hierarchical mixing angles, $ \theta_1 \gg \theta_2 \gg \theta_3$, in quark sector.  In the case of solar neutrino oscillation, $\overline{P}(\nu_{1L} \rightarrow \nu_{1L})_{\mbox{eff}}$ is obtainable from the time evolution governed by the effective Hamiltonian Eq.(28), which is very similar to that in the MSW mechanism with maximal mixing angle, $\pi /4$ . To be precise, in the Hamiltonian (28) the matter effect is not $\sqrt{2} \ G_{F} \ N_{e} $ as in the case of MSW. The difference, however, is not large, as long as the generation mixings are small and also because the contribution of neutral current, 
the term proportional to $N_{n}$,  is relatively suppressed compared with that of charged current by a factor $\sim 1/12$.  On the other hand, very  recently we have heard of the news \cite{Kamsolar} that the data of super-Kamiokande on solar neutrinos favors the MSW-type solution with large mixing angle. Thus our scenario of pseudo-Dirac provides a natural framework to derive the large angle 
solution suggested by the data. 
 In the case of atmospheric neutrino oscillation, 
the factor in front of  $\mbox{sin}^2 (\frac{m_2 |\epsilon_2|}{2E} t)$ is $1 = \mbox{sin}^2 (2 \times \frac{\pi}{4})$, which just corresponds to a vacuum 
oscillation with maximal mixing, strongly suggested by the data on atmospheric neutrinos \cite{Kamatm}. 

In a contrary, in the case of LSND our formula gives that of ordinary 2 generation scheme with small generation mixing, if we identify $4|U_{e 2}|^2$ with $\mbox{sin}^2 2\theta_1$. This is just consistent with the experimental data \cite{LSND}, which says $\mbox{sin}^2 2\theta_1 \leq 0.04$, when combined with the data  from BUGEY experiment. Another meaningful constraint on the generation mixing may come from  the data of CHOOZ experiment \cite{Chooz}. As the mass-squared difference which is sensitive to the CHOOZ experiment is comparable to that in atmospheric neutrino oscillation, we may write down a similar formula to Eq.(20) for the disappearance of $\overline{\nu}_e$, 
\begin{eqnarray}
1 - P(\nu_{e L} \rightarrow \nu_{e L})_{\mbox{CHOOZ}} &=&  1 - |U_{e 1}|^4 - |U_{e 2}|^4 
\mbox{cos}^2 (\frac{m_2 |\epsilon_2|}{2E} t) - \frac{1}{2} |U_{e 3}|^4 \nonumber \\ 
&\simeq& 2 |U_{e 2}|^2 ,   
\end{eqnarray}
where the unitarity of $U$ matrix was used and an approximation of $|U_{e3}| \ll |U_{e2}|$ was useful to simplify the result. The upper bound on the probability of disappearance from CHOOZ puts a bound 
\begin{equation}
|U_{e 2}|^2 \sim \mbox{sin}^2 \ \theta_{1} \ \leq 0.05, 
\end{equation}
which is a little weaker than the upper bound stated above. 

\noindent 4.2 \ Problems to be settled 

We have seen that the pseudo-Dirac scenario just provides the favored solutions to the solar and the atmospheric neutrino problems with (almost) maximal mixings, suggested by the recent data, invoking the oscillations mainly into sterile states, while LSND data is naturally explained by ordinary generation mixing between active states with  a small mixing. We naively expect that neutrino oscillations of solar or atmospheric neutrino into an active state and a sterile state cannot be clearly discriminated, since basically these experiments are disappearance experiments. Roughly speaking this expectation is certainly true, but the data from Super-Kamiokande experiment have reached the precision which is enough to distinguish these two cases. The most recent data seem to regard the maximal mixing solutions of neutrino oscillations into sterile states with disfavour \cite{Kamsolar},  \cite{Kamatm2}. 
We would like to discuss some possible ways to avoid the difficulty for each case. 

\noindent Atmospheric neutrino 

    Though we have neglected the matter effect in the atmospheric neutrino 
oscillation, the matter effect of the Earth becomes non-negligible for higher neutrino energies. It has been pointed out that $\nu_{\mu}$ oscillations into $\nu_{\tau}$ and  a sterile state have different zenith-angle dependence, as only in the case of the oscillation into the sterile state the matter effect affects the time-evolution of the neutrino states. Comparing with the data, combining the analysis of neutral current enriched events, the Super-Kamiokande collaboration claims that the oscillation into the sterile state with maximal mixing is regarded with disfavour \cite{Kamatm2}. 
The possible ways to evade this problem, we can think of, are the following. \\ 
a. \  When the oscillation to the sterile state is analyzed, simplified 2 states system of  $(\nu_{\mu}, \nu_{S})$ \ ($\nu_{S}$ denoting a sterile state) is assumed. In the scenario of pseudo-Dirac, however, we have 6 neutrino states to participate in the oscillation, and the formula for the oscillation, as seen in Eq.(20), is different from that in the simple 2 states system, typically having additional constant terms (for non-vanishing generation mixings). Including the matter effect of the Earth, our formula is modified into 
\begin{equation}
\overline{P}(\nu_{\mu L} \rightarrow \nu_{\mu L})_{\mbox{atm}} = |U_{\mu 2}|^4 \ \overline{P}(\nu_{2L} \rightarrow \nu_{2L})_{\mbox{eff}} +  \ |U_{\mu 1}|^4
                      + \frac{1}{2} \ |U_{\mu 3}|^4,   
\end{equation}
where the survival probability $ \overline{P}(\nu_{2L} \rightarrow \nu_{2L})_{\mbox{eff}}$ in the effective 2 states system $(\nu_{2L}, \ \overline{\nu}_{2L})$ is that for the mass-squared difference $2 \ m_2 \epsilon_{2}$ and the matter effect $ \sum_{\alpha = e}^{\tau} |U_{\alpha 2}|^2 a_{\alpha}$.  Thus both the depletion rate of atmospheric $\nu_{\mu}$ and the zenith angle dependence should be reanalyzed by use of this formula before some definite conclusion is derived.  \\ 
b. \  We just would like to point out that there is a claim that $\nu_{\mu} \rightarrow \nu_{S}$ oscillation with almost maximal mixing may not be ruled out, even in the simple 2 states system of  ($\nu_{\mu}$, $\nu_{S}$). It has been pointed out that a $\chi^2$ analysis of the recent data does not exclude the maximal  $\nu_{\mu} \rightarrow \nu_{S}$ oscillation solution with any significant confidence level, once various theoretical uncertainties and experimental systematic errors are included \cite{Foot}.  

\noindent Solar neutrino 

The Super-Kamiokande collaboration claims that the solar neutrino oscillation into the sterile state with maximal mixing is not favoured \cite{Kamsolar}.  The point is that $\nu_{e} \rightarrow \nu_{\mu}, \nu_{\tau}$ and $\nu_{e} \rightarrow \nu_{S}$ oscillations give slightly different contributions in the Super-Kamiokande detector, as the final active states contribute to the event rate, while the sterile state does not. 
Accordingly, the survival probability of solar neutrino should be relatively 
higher in the case of oscillation into the sterile state. Thus the ``Be neutrino problem" in the chlorine experiment becomes more severe in the 
sterile case. The possible ways out of this problem, we can think of, are the following. \\ 
a. \   Again the claim that the solar neutrino oscillation to the sterile state is not favoured is based on the analysis assuming a simplified 2 states system of ($\nu_{e}, \nu_{S}$). In the scenario of pseudo-Dirac, however, we have 6 neutrino states to participate in the oscillation, and the formula Eq.(29) should be utilized to see whether the oscillation into the sterile state can accommodate all data of solar neutrino experiments or not. \\ 
b. \   In the case of solar neutrino oscillation, the presence of solar magnetic field is potentially important, though we have ignored it in the above 
discussions. It has been pointed out that in the presence of the magnetic field, MSW type oscillation $\nu_{eL} \rightarrow \overline{\nu}_{eL}$ may be followed by a RSFP type oscillation $\overline{\nu}_{eL} \rightarrow \overline{\nu}_{eR}$, ignoring generation mixing \cite{Kobayashi}. The final state, now being an active state, contributes to the event rate and may remedy the  ``Be neutrino problem". \\ 
c. \   In the above argument, Majorana masses have been treated as  small perturbation, but it may not be unnatural to expect that in the first generation the Dirac mass is so small that the effect of Majorana masses is 
relatively enhanced, leading to a relatively small mixing between the active and sterile states.  

\noindent Cosmological problem 

There is another type of problem, i.e. cosmological problem. 
The well-known limit of number of effective neutrino species during nucleosynthesis puts a stringent bounds on the mass-squared differences and mixings of neutrino oscillations into sterile states \cite{Barbieri}. It, however, has been pointed out that once relatively large relic neutrino asymmetry $L_{\nu}$, say 
$L_{\nu} \geq 10^{-4}$, is realized such problem can be evaded. For the details  of the argument refer to  \cite{Foot2}.

We finally briefly comment on the related issue, i.e. neutrino-less double $\beta$-decay. In our pseudo-Dirac scenario, the relevant lepton number violating Majorana mass is given as 
\begin{equation} 
\frac{1}{2} \ (U_{\alpha i})^{2} \ (m_{iS} - m_{iA}) \simeq \frac{1}{2} \ 
(U_{\alpha i})^{2} \ |\epsilon_{i}|, 
\end{equation} 
which is well below the experimental upper bound under the mass hierarchy (17),  (18) with $\epsilon_{i} \ll m_{i}$ and for small generation mixings. This is basically because the pseudo-Dirac neutrinos are almost Dirac particles and the lepton number is only slightly violated by their masses. 

\vspace{0.5 cm}
\leftline{\bf 5. Theoretical framework for the pseudo-Dirac neutrinos}
\vspace{0.2 cm}

Though there have already appeared a few attempts to construct models for the 
pseudo-Dirac neutrinos \cite{Chang}, in this article we instead list up the problems to be resolved before a realistic model is constructed, and argue about possible theoretical framework to provide natural mechanisms to solve the problems. 

First of all it is worth noticing that the smallness of Majorana masses needed to realize the pseudo-Dirac neutrinos satisfies the naturalness condition of 't Hooft \cite{'t Hooft}, since if Majorana masses are absent, $M_L = M_R = 0$, the symmetry of the theory, i.e. the lepton number symmetry is enhanced. So the smallness will be stable under the radiative correction. 

We, however, still have the following problems to be settled at the classical level: \\ 
1. How to explain the relation $M_R \ll M_D$ ? \\ 
2. How to explain the relation $M_L \ll M_D$ ? \\
3. How to explain the smallness of $M_D$ itself ? \\ 
We will discuss the possible theoretical frameworks to resolve these problems 
successively below.      

\noindent 1. Problem 1 

\noindent a. \ 4-dimensional framework 
 
In considering the possible theoretical framework to resolve this problem, it 
may be helpful to reconsider the conventional see-saw mechanism in the language of gauge invariant operators. Suppose that the gauge symmetry of our world is U(1)$_{em}$ (QED), then the Majorana mass term $m_L \ \nu_{L}^2$ is gauge invariant. Therefore there will 
be no reason to expect that $m_L$ should be small. As the matter of 
fact, the gauge symmetry of the standard model does not allow the mass 
operator, as it is gauge variant. Thus the Majorana mass is provided by an irrelevant operator $\frac{1}{M} \ L_{L}^2 \ H^2$, where $H$ denotes the Higgs doublet and $L_{L} = (\nu_{L}, l_{L})^{t}$ is a lepton doublet. As $M$ is a gauge invariant mass, it can be arbitrarily large. Thus the essence of see-saw mechanism may be understood as the decoupling of some gauge 
singlet heavy particle with mass $M$ (which need not to be $\nu_R$). It, therefore, will be not unnatural to expect that a similar thing 
happens for the right-handed Majorana masses, as well. The Majorana mass term $m_R \ \nu_{R}^2$ is gauge invariant in the standard model 
and $m_R$ is regarded to be quite large. This may not be true in some physics beyond the standard model. For instance in the left-right symmetric model SU(2)$_L$ $\times$ SU(2)$_R$ $\times$ U(1) \cite{Mohapatra} with a SU(2)$_R$ doublet Higgs $H_R$ (in addition to the ordinary SU(2)$_L$ doublet Higgs), the  $m_R \ \nu_{R}^2$ operator is no longer gauge invariant, and 
will be replaced by an irrelevant operator $\frac{1}{M'} \ L_{R}^2 \ H_{R}^2$, 
with  $L_{R} = (\nu_{R}, l_{R})^{t}$. 
Thus the decoupling of a gauge singlet heavy particle whose mass $M'$ is much larger than the scale of $\mbox{SU}(2)_{R}$ breaking, $M' \gg \langle H_{R} \rangle$, may imply the smallness of $m_R$.  The heavy particle can be identified with a gauge singlet fermion $S$, having a Yukawa coupling $L_{R} \ S \ H_{R}$ and a large Majorana mass $M'$. A diagram with the exchange of $S$ yields the irrelevant operator $L_{R}^2 \ H_{R}^2$. 

\noindent b. \ Framework with extra dimensions 

Recently there has appeared revived interest in higher dimensional theories 
with extra dimensions as a possible solution to hierarchy problem \cite{Arkani-Hamed}, \cite{Hatanaka}, \cite{Randall}. In the scenario of large extra dimension \cite{Arkani-Hamed}, the higher dimensional Planck scale is regarded as comparable to the weak scale, while in the scenario of small extra dimension \cite{Randall} all masses in the visible brane are claimed to be strongly suppressed by the ``warp factor". Therefore the conventional see-saw mechanism \cite{Seesaw}, which needs large right-handed Majorana masses ($m_{R} \gg M_{W}$),  may not 
work. In the case of pseudo-Dirac scenario, on the contrary, what we need is very small or even vanishing $m_{R}$. It is interesting to note that the presence of extra-dimensions may provide a natural mechanism to realize this. For instance,  let us consider a theory in 5-dimensional space-time, where the  particles of the standard model , including $\nu_{L}$,  are assumed to reside on a  3-brane, while gauge singlet fields, such as $\nu_{R}$, may reside in the bulk \cite{Arkani-Hamed2}, \cite{Smirnov}.  We note that in 5-dimensional space-time  Majorana spinor is known not to exist. (Majorana spinors exist only in the space-time of $D = 2,3,4 \ (\mbox{mod} \ 8)$.)  Thus the mass term $m_{R} \ \nu_{R}^{2} + h.c.$ is not Lorentz invariant in this space-time and we have vanishing  $m_{R}$. 
Actually in such odd dimensional space-time chiral fermions do not exist, and 
$\nu_{R}$ should be accompanied by a gauge singlet $\tilde{\nu}_{L}$ to form a 
full spinor $\psi = (\nu_{R}, \tilde{\nu}_{L})^{t}$. The full spinor may have a large Dirac mass and $\nu_{R}$ may be decoupled from low energy phenomena. This difficulty may be evaded when the extra space is an orbifold $S^{1}/Z_{2}$, the extra-space suggested by the recent works \cite{Arkani-Hamed}, \cite{Randall}, \cite{Horava}. This is because the  discrete symmetry $Z_{2}$ just corresponds to a symmetry under the transformation $\psi \rightarrow \gamma_{5} \psi$, which in turn behaves as a chiral transformation in the 4-dimensional sense, thus making the Dirac mass term prohibited. Strictly speaking, $\psi$ may also have masses due to non-zero K-K modes. For relatively small sizes of the extra space, such non-zero modes tend to be decoupled from the system and do not significantly affect the above argument. 

\noindent 2. Problem 2 

  In some sense, this problem may not be a real challenge; in any viable 
model the relation  $M_L \ll M_D$ must be automatically built in, since 
otherwise custodial symmetry is significantly violated by the VEV of a SU(2) 
triplet representation and  $\Delta \rho = \rho - 1$ gets sizable contribution, in contradiction with the data. Furthermore if we have already got some mechanism to realize small but non-vanishing $m_{R}$ in a mechanism discussed above, we 
are satisfied with vanishing $m_{L}$ just as in the Standard Model. In a model with vanishing $m_{R}$, however, it becomes crucial for the pseudo-Dirac scenario to slightly violate the lepton number  by small but non-vanishing $m_{L}$. 
It is worth noticing that even in a higher dimensional model with brane picture  the $\nu_{L}$ is allowed to live only on the brane, and the issue concerning 
$m_{L}$ is essentially 4-dimensional. 

If we wish to get the small $m_{L}$ at the classical level as the form of a renormalizable operator we should introduce a SU(2)$_{L}$ triplet Higgs $H_{T}$, whose VEV should be small not to contradict with the custordial symmetry. Then a marginal operator $L_{L}^{2} H_{T}$ gives small $m_{L}$. Or we may invoke an irrelevant operator $\frac{1}{M} \ L_{L}^2 \ H^2$, as we discussed above, with  a large mass scale $M \gg \langle H \rangle$. The operator may be the result of the exchange of a hypothetical gauge singlet $S'$, which has a Yukawa coupling 
$L_{L} S' H$ and a large Majorana mass $M$. The S' should not be identified with $\nu_{R}$. If we work in the framework of extra dimension with a brane, however,  there may not be a good reason to assume that the gauge singlet $S'$ resides only on the brane. Even if $m_{L}$ is forbidden at the classical level, the small $m_{L}$ may still be produced at the loop level, as far as there is some seed to violate the lepton number. The prototype model of this kind may be the model where the presence of a charged SU(2) singlet scalar violates the lepton number explicitly \cite{Zee}.

\noindent 3. Problem 3 

In pseudo-Dirac scenario,  Dirac masses provide mean masses of neutrinos. Thus  crucial problem is how to explain the smallness of the neutrino Dirac masses 
compared with those of charged leptons or quarks. 

\noindent a. \ 4-dimensional framework 
 
In the 4-dimensional framework ``Dirac see-saw" mechanism has been put forwarded \cite{Chang}, in which the marginal operator to give Dirac mass term $H \ \overline{\nu}_L \ \nu_{R}$ is forbidden, by a discrete symmetry, and the Dirac masses are provided by an irrelevant operator with $d > 4$, just as happens 
in the conventional see-saw mechanism. Thus if the coefficient is sufficiently suppressed by the inverse of some large mass scale, we get small Dirac masses. 
 
\noindent b. \ Framework with extra dimensions 

Another intriguing possibility to realize the small neutrino Dirac masses is to invoke the presence of extra space \cite{Arkani-Hamed2}, \cite{Smirnov}. For instance in the $4+n$-dimensional theory with a 3-brane, the original higher-dimensional Yukawa coupling  $f_{0} \ H \ \overline{\nu}_L \ \nu_{R}$ has a Yukawa coupling constant $f_{0}$, which behaves as $1/\sqrt{M_{f}^{n}}$, with $M_{f}$ being fundamental mass scale of the theory. The 4-dimensional Yukawa coupling 
$f_{4}$ is thus given by $f_{4} \sim 1/\sqrt{M_{f}^{n} V}$ ($V$ : the volume of the extra dimension). The factor $1/\sqrt{V}$ comes from the overlap of the three fields on the brane, and $f_{4}$ may be suppressed by the largeness of the extra dimension. In fact in the ADD model \cite{Arkani-Hamed}, by use of the relation $M_{pl}^{2} = M_{f}^{2+n} V$,  we get a $n$-independent result 
\begin{equation} 
f_{4} \sim \frac{M_{f}}{M_{pl}},   
\end{equation} 
which is $\sim 10^{-16}$ for e.g. $M_{f} \sim 1 TeV$.  In this way, a small Dirac mass, $\sim f_{4} \ M_{W} \sim f_{4} \ M_{f}$, is achievable.

\noindent {\bf Acknowledgment}

The authors would like to thank W.J. Marciano, Y. Okada, O.L.G. Peres, and O. Yasuda  for useful and informative discussions. One of the authors (C.S.L.) would like to thank the members of the BNL theory group for their hospitality where a  part of this work was completed. This research was supported in part by the Grant-in-Aid for Scientific Research of the Ministry of Education, Science and Culture, No.09246105, No.12047219, No.12640275.

\end{document}